\documentclass[aps,prd,twocolumn,showpacs,preprintnumbers,amsmath,amssymb,floatfix]{revtex4}

\usepackage{graphicx}
\usepackage{dcolumn}
\usepackage{bm}
\usepackage{psfrag}

\newcommand\be{\begin{equation}}
\newcommand\ee{\end{equation}}
\newcommand\bea{\begin{eqnarray}}
\newcommand\eea{\end{eqnarray}}
\newcommand\ba{\begin{array}}
\newcommand\ea{\end{array}}

\begin{document}

\preprint{SACLAY-T11-032}

\title{Large Gravitational Wave Background Signals in Electroweak Baryogenesis Scenarios}

\author{Jos\'e M. No$^{1}$}

\affiliation{$^{1}${\it Institut de Physique Th\'eorique, CEA/Saclay, F-91191 
Gif-sur-Yvette C\'edex, France}}

\date{\today}

\begin{abstract}
The bubble wall velocity in an electroweak first order phase transition is a key quantity both for electroweak baryogenesis and for 
the production of a stochastic background of gravitational waves that may be probed
in the future through gravitational wave experiments like LISA or BBO. We show that, contrary to the conclusion drawn from previous studies, 
it is actually possible to generate a potentially large gravitational wave signal while satisfying the requirements for 
viable electroweak baryogenesis, once the effects of the hydrodynamics of bubble growth are taken into account. Then, the observation of 
a large gravitational wave background from the electroweak phase transition would not necessarily rule out electroweak baryogenesis 
as the mechanism having generated the observed baryon asymmetry of the universe. 
\end{abstract}

\maketitle

\section{Introduction}

The study of bubble growth in cosmological first order phase transitions 
is very relevant for many phenomena having possibly occured in the early universe 
such as electroweak baryogenesis \cite{Cohen:1990py,Joyce:1994fu} 
or the production of a stochastic 
background of gravitational waves \cite{Witten:1984rs,Kosowsky:1991ua,Kamionkowski:1993fg}. 
A first order phase transition proceeds by bubble nucleation
and expansion, and the velocity of the expanding bubble walls plays an essential role 
both in electroweak baryogenesis and in gravitational wave production, since the efficiency of 
both processes strongly depends on its value. 

Treatments of the bubble wall velocity generally assume that friction from 
the plasma balances the initial pressure difference that drives the bubble 
expansion, so that the wall reaches a constant speed after a short period of 
acceleration. Assuming that the free energy of the Higgs field is released into the plasma, a hydrodynamic
treatment of the plasma can be used to determine the fluid motion
\cite{Landau, Steinhardt:1981ct}, but this approach leaves the wall velocity 
as a free parameter as long as the microscopic mechanism of friction is unknown. 
Ultimately, the wall velocity can be fixed using the equation of motion for the Higgs field,
that takes into account the friction of the plasma \cite{Moore:1995si,Ignatius:1993qn}. 

However, successful electroweak baryogenesis and sizable gravitational wave production require very different wall velocities.
The electroweak baryogenesis mechanism is based on the 
interaction between the expanding bubble wall
and the plasma in front of it, leading to a CP asymmetric reflection on the wall of certain particle species, 
and the subsequent diffusion of these  
particle asymmetries into the plasma in front of the bubble wall \cite{Joyce:1994fu}, where sphalerons are active and capable
of converting the CP asymmetry into a net baryon number. Then the generated baryon number is carried into the broken 
phase as the wall passes by (where it stays frozen if the sphaleron processes are sufficiently suppressed in the broken phase). 
In order for this whole mechanism to be 
effective the diffusion timescale has to be smaller than the time the wall takes to sweep through the 
plasma just in front (otherwise diffusion is ineffective and the generation of baryon number is strongly supressed), and 
this puts an upper bound on the relative velocity between 
the wall and the plasma in front $V \lesssim D/L_{w} \sim 0.15 - 0.3$ (being $D$ a certain diffusion constant and $L_{w}$ 
the wall thickness) \cite{Joyce:1994fu}. Moreover, it is generally stated 
that subsonic wall velocities are always needed, because effective diffusion cannot take place for supersonic walls. 
On the other hand, fast moving walls are essential for the production of a sizable amount of gravitational radiation 
in bubble collisions \cite{Kosowsky:1991ua, Kamionkowski:1993fg, Caprini:2007xq, Huber:2008hg, Caprini:2009fx},
or turbulence in the plasma \cite{Kosowsky:2001xp,Caprini:2006jb}. 
In particular, for bubble collisions the gravitational wave amplitude 
is roughly proportional to $V_w^3$ \cite{Caprini:2007xq, Huber:2008hg}, with an extra implicit 
$V_w$-dependence through the efficiency coefficient $\kappa$ for transforming the available 
energy from the phase transition into plasma bulk motion (which is in turn responsible for 
the generation of the gravitational wave background during the bubble 
collisions \cite{Kamionkowski:1993fg,EKNS}), and this dependence further suppresses 
the gravitational wave signal for small wall velocities (this suppression also affects the 
gravitational wave signal generated from turbulence). All this has established the common lore that both phenomena 
cannot happen in the same scenario.

In the Standard Model, and for values of the Higgs mass above the LEP bound $M_h > 114.4$ GeV \cite{LEP}, 
the electroweak phase transition is found not to be 
of first order, but rather a smooth cross-over \cite{Kajantie}. However, there are many possible 
theories beyond the Standard Model in which the electroweak phase transition may naturally be of first order,
such as extensions of the MSSM \cite{Pietroni:1992in,KonstandinNMSSM, Blum} 
(in the MSSM itself, the region of parameter space where a first order phase
transition leading to electroweak baryogenesis is achieved is currently very tightly constrained \cite{Germano}),
Two-Higgs-Doublet models \cite{Bochkarev:1990fx,Turok:1991uc,Davies:1994id,Cline:1996mga}, 
singlet field extensions of the Standard Model \cite{AndersonHall,Ham:2004cf,Espinosa:2007qk,Profumo:2007wc,EKNQ}, 
composite Higgs models and others.
Also, in \cite{Servant} the electroweak phase transition was studied for the Standard Model considered as an effective theory
with a low cut-off, finding that the inclusion of higher dimensional operators in the Higgs potential may give rise to
a rather strong first order phase transition. As it has been discussed above, for these or any other model leading
to a first order electroweak phase transition, the value of the velocity of the expanding bubbles is a key parameter
for the study of both electroweak baryogenesis and gravitational wave production at the phase transition. 
For the case of the MSSM the value of $V_w$ was found to be quite small over all
the available parameter space \cite{Friction}, but it is expected that it may be much larger in many of the models discussed above,
since the electroweak phase transition in those cases is much more strongly first order, since the wall velocity
increases with the strength of the phase transition \cite{Ignatius:1993qn}.      

Here we will show that in contrast to the case of gravitational wave production, where the relevant velocity
is indeed the speed of the wall $V_w$, in electroweak baryogenesis the relevant velocity (being the relative velocity between 
the bubble wall and the plasma just in front the wall $v_+$) is in general lower than $V_w$, and this effect becomes more important 
as the phase transition gets stronger. Then, it is possible to have a sizable gravitational wave production
(through a relatively large $V_w$, a natural possibility in many of the scenarios beyond the Standard Model mentioned
above, where the electroweak phase transition can be strongly first order) while the electroweak baryogenesis mechanism is still effective 
($v_+$ is sufficiently low). 

This article will be organized as follows: In section 2 a summary of the hydrodynamic analysis 
of bubble expansion is given. Then, in section 3 the use of $v_+ $ instead 
of $V_w $ as relevant velocity for electroweak baryogenesis is motivated, and in section 4 it is shown that
$v_+ < V_w$ (and $v_+ \ll V_w$ for strong transitions). In section 5 the gravitational wave amplitude in viable electroweak
baryogenesis scenarios is obtained, and we conclude in section 6.

\section{Hydrodynamic Relations.}

The hydrodynamic analysis of the system consisting 
on the Bubble wall and the plasma surrounding it \cite{Landau, Steinhardt:1981ct, EKNS} rests on two basic
assumptions: energy-momentum conservation in the system and local thermal equilibrium (LTE) in the plasma. 
This last assumption is reasonable since local equilibration due to the interactions between the 
particle species in the thermal plasma is much faster than the actual macroscopic movement of the 
plasma. Local thermal equilibrium is also crucial for the consistency of the fluid approximation 
applied in electroweak baryogenesis computations (see \cite{Joyce:1994fu}). 

The energy-momentum tensor of the Higgs field $\phi $ is given by 

\be
T_{\mu\nu}^{\phi} = \partial_{\mu}\phi \partial_{\nu}\phi -g_{\mu\nu} 
\left[ \frac{1}{2} \partial_{\rho}\phi \partial^{\rho}\phi - V_0 (\phi)\right] \, ,
\ee

where $V_0 (\phi)$
is the renormalized vacuum potential. If the
plasma is locally in equilibrium its energy-momentum tensor can be parametrized as

\be
T_{\mu\nu}^{plasma} = w \, u_\mu u_\nu  - g_{\mu\nu} \, p \, ,
\ee

where $w$ and $p$ are 
the plasma enthalpy and pressure, respectively. The quantity $u_\mu$ is the four-velocity 
field of the plasma, related to the three-velocity $\mathbf{v}$ by $u_\mu = (\gamma, \gamma\mathbf{v} )$.
The enthalpy $w$, the entropy density
$\sigma$ and the energy density $e$ are defined by

\be
w \equiv T\frac{\partial p}{\partial T}\ ,\quad
\sigma \equiv \frac{\partial p}{\partial T}\ ,\quad
e\equiv T\frac{\partial p}{\partial T} -p\ ,
\ee
where $T$ is the temperature of the plasma. Conservation of energy-momentum is given by 
$\partial^\mu T_{\mu\nu}^{\phi} + \partial^\mu T_{\mu\nu}^{\mathrm{plasma}} = 0$. 
Since we are interested in a system where the bubble expands at a constant speed, energy-momentum
conservation reads in the wall frame (assuming no time dependence, and with the wall and fluid 
velocities aligned in the $z$ direction) $\partial_z T^{zz} = \partial_z T^{z0} = 0$. Integrating 
these equations across the wall and denoting the phases in front and behind by subscripts $+$
(symmetric phase) and $-$ (broken phase) one obtains the matching equations in the wall frame:
\bea
\label{eq:wall_constr}
w_+ v^2_+ \gamma^2_+ + p_+  = w_- v^2_- \gamma^2_- + p_-  \nonumber \\
w_+ v_+ \gamma^2_+ = w_- v_- \gamma^2_- \quad \quad
\eea

From these equations we can obtain the relations \cite{Steinhardt:1981ct}
\be
\label{eq:vvs0}
v_+ v_- = \frac{p_+  - p_-}{e_+ - e_-}\ , \quad \quad
\frac{v_+}{ v_-} = \frac{e_- + p_+ }{e_+ + p_-}\ . 
\ee
In order to proceed further, one needs to know the equation of state (EoS) 
for the plasma. A parametrization that accounts for deviations from the so-called bag EoS
(usually used in hydrodynamical studies of phase transitions in the early universe 
\cite{Steinhardt:1981ct,Kamionkowski:1993fg}) is (see \cite{EKNS})
\be
\label{eosbag1}
p_+ = \frac{1}{3} a_+ T_+^4 \quad \quad
e_+ = a_+ T_+^4 
\ee
\bea
\label{eosbag2}
p_- = \frac{1}{3}a_- T_-^4 + \epsilon \equiv a_+ T_+^4 \left(\frac{1}{3 r} + \alpha_+\right) \nonumber \\ 
e_- = a_- T_-^4 - \epsilon \equiv a_+ T_+^4 \left(\frac{1}{r} - \alpha_+\right) 
\eea

where we have defined $\alpha_+\equiv\frac{\epsilon}{a_+ T_+^4}$ and $r\equiv\frac{a_+T_+^4}{a_-T_-^4}$. 
The quantity $\alpha_+$ is approximately the ratio of vacuum energy difference to thermal energy in front of the wall, 
and characterizes the strength of the phase 
transition (the larger $\alpha_+$ the stronger the transition), 
and $a_\pm$ are related to the number of relativistic d.o.f
in the symmetric and broken phases. Using (\ref{eosbag1}) and (\ref{eosbag2}) we 
can write the relations (\ref{eq:vvs0}) as
\be
\label{eq:vvs}
v_+ v_- = \frac{1 - (1-3\alpha_+) r }
{3 - 3( 1 + \alpha_+ )r} \ , \, \, \quad 
\frac{v_+}{ v_-} = \frac{3  + (1-3\alpha_+) r}
{1 + 3(1 + \alpha_+)r}
\ee

The two equations (\ref{eq:vvs}) can be combined to give
\begin{small}
\bea
\label{eq:vvs2}
v_+ = \frac{1}{1+\alpha_+}\left[\left(\frac{v_-}{2}+\frac{1}{6
v_-}\right) \right. \quad \quad \quad \quad \quad \quad \quad \nonumber \\
\left. \pm \sqrt{\left(\frac{v_-}{2}+\frac{1}{6 v_-}\right)^2 +
\alpha_+^2 +\frac{2}{3}\alpha_+}-\frac{1}{3} \right]
\eea
\end{small}
so that there are two branches of solutions, corresponding to the
$\pm $ signs in (\ref{eq:vvs2}). 

In a concrete model where $p = -\mathcal{F}$ (the free energy or finite-temperature effective potential) 
is known, the thermodynamic potentials can be calculated in the two phases and the temperature $T_N$ at 
which the phase transition happens is determined using standard
techniques \cite{cite:fate}. Still, there are three unknown quantities
($T_-$, $v_+$ and $v_-$) and two equations (\ref{eq:vvs0}), so
up to this point all hydrodynamically viable solutions are
parametrized by one parameter, usually chosen to be the wall velocity $V_w$.

Next, we briefly review the solutions of the plasma
velocity $v$ \cite{Laine:1993ey,EKNS}. Applying energy-momentum 
conservation in the plasma $\partial^\mu T_{\mu\nu}^{\mathrm{plasma}} = 0$ (far from the wall
$T_{\mu\nu}^{\phi}$ just gives a constant background that plays no role in the energy-momentum conservation), 
we obtain a set of hydrodynamic equations. Since there is no intrinsic macroscopic length scale
in the system, solutions to these equations are
self-similar and only depend on $\xi = r/t$, with $r$ being
the radial coordinate of the bubble and $t$ the time since
nucleation. The plasma then fulfills the equations
\bea
\label{eq:v_diff}
2 \frac{v}{\xi} = \gamma^2 (1 - v \xi) 
\left[ \frac{\mu^2}{c_s^2} -1 \right] \partial_\xi v  \nonumber \\
\frac{\partial_\xi w}{w} =4 \gamma^2 \, \mu(\xi,v) \partial_\xi v
\eea
where $c_s=1/\sqrt{3}$ denotes the velocity of sound in the plasma and
$\mu(\xi, v)$ is the Lorentz-transformed fluid velocity $\mu(\xi, v) = \frac{\xi - v }{1 -\xi v}$.
Generally, there are three different types of
solutions to (\ref{eq:v_diff}) with the boundary conditions (\ref{eq:vvs2}) \cite{EKNS, Laine:1993ey}: 
detonations, deflagrations and hybrid
solutions. In detonations the bubble wall expands at supersonic velocities
and the vacuum energy of the Higgs leads to a rarefaction wave behind
the wall, while the plasma in front is at rest. In
this case, the wall velocity is $V_w = v_+ > v_-$, and therefore 
detonations are identified with the $+$ branch of solutions in (\ref{eq:vvs2}). In
deflagrations, the plasma is mostly affected by reflection of
particles at the bubble wall and a compression wave builds up in front
of the wall while the plasma behind is at rest. In this case,
the wall velocity is identified with $V_w = v_- > v_+$, corresponding to the $-$ 
branch of solutions in (\ref{eq:vvs2}). While ``pure" deflagrations are subsonic, 
the hybrid case occurs for supersonic deflagrations where both effects (compression 
and rarefaction wave) are present. From now on we will focus on deflagrations and hybrids
since for them $V_w > v_+$. Also, in this case $T_+ > T_N $, and so $\alpha_+ < \alpha_N \equiv\frac{\epsilon}{a_+ T_N^4}$
(being $\alpha_N$ the actual measure of the strenght of the phase transition). 

\section{$v_{+}$ vs $V_{w}$ in Electroweak Baryogenesis.}

\begin{figure}[ht]
\begin{center}
\includegraphics[width=0.45\textwidth, clip]{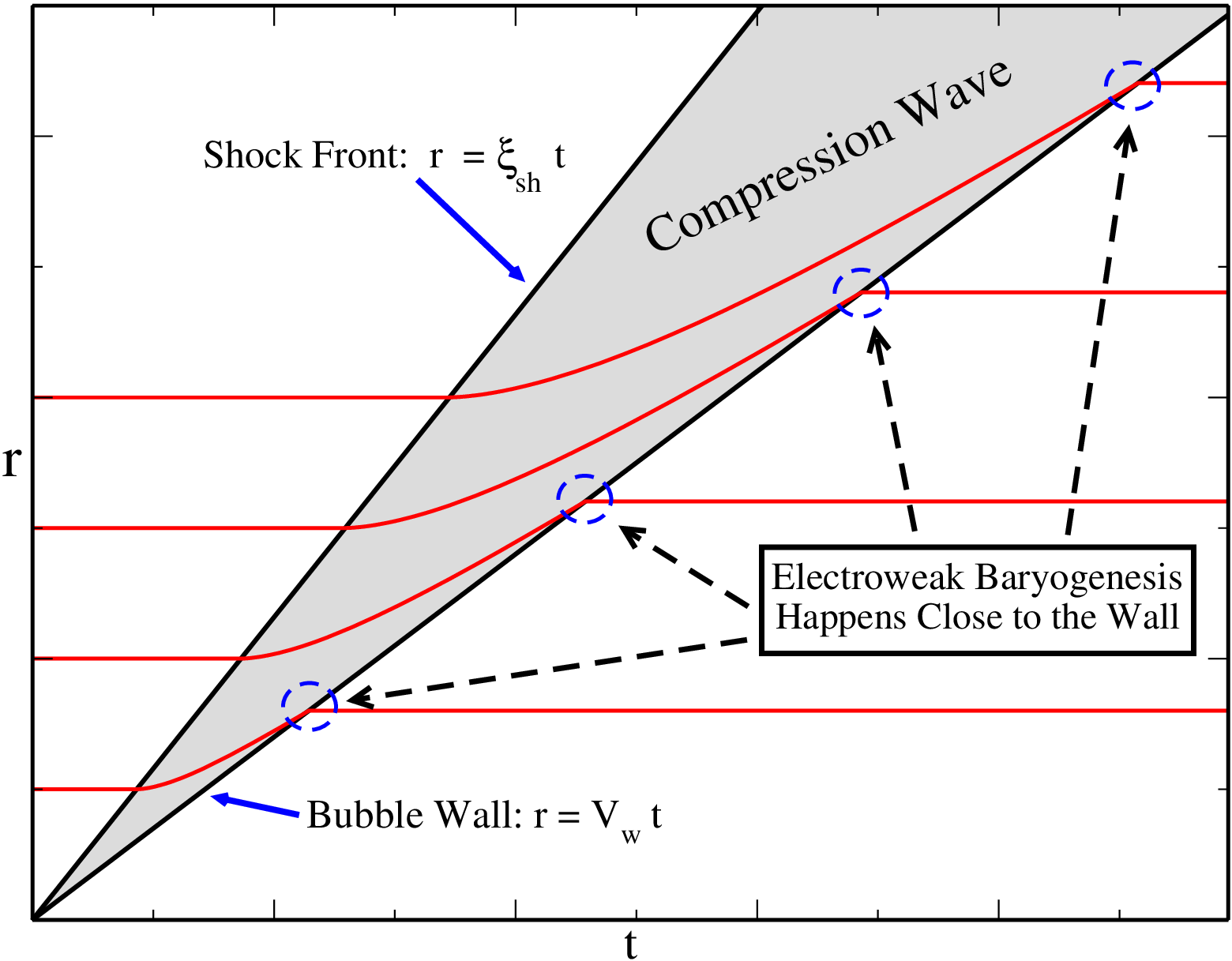}
\caption{\label{fig:1}
\small Movement of plasma volume elements with time.}
\end{center}
\end{figure}

Consider the evolution of a plasma volume element 
as the compression front and bubble wall reach it and pass by, in the case of a deflagration (Figure \ref{fig:1}). The element
is initially at rest (its position $r = \mathrm{cte}$), and as it enters the compression wave it acquires a velocity, that
grows until reaching $v(V_w) = \frac{V_w -v_+}{1- V_w v_+}$ close to the bubble wall. Then, due to the 
compression wave, the relative velocity between the wall and the volume element just in front of it 
is $v_+$ independently from the details of the electroweak baryogenesis mechanism close to the wall, 
just relying on LTE applying all over the compression wave. In the electroweak baryogenesis analysis 
the relevant velocity is the perturbation $\overline{V}$ with respect 
to the background velocity of the plasma $v_+$. However, in \cite{Joyce:1994fu} (and subsequent electroweak baryogenesis studies) 
the background velocity 
of the plasma was taken to be $V_w$, neglecting hydrodynamics. We find that in the presence of a compression wave
and in the wall reference frame, the velocity of the plasma should be $V = - v_+ + \overline{V}$ 
instead of $V = - V_w + \overline{V}$, 
and so the dependence of the electroweak baryogenesis mechanism on the wall velocity 
$V_w$ extracted from \cite{Joyce:1994fu} is actually a dependence on $v_+$. The same argument applies to the upper 
bound on $V_w$ from \cite{Joyce:1994fu} from the requirement of effective diffusion of the CP 
asymmetric current in front of the wall. The bound should now be regarded as 
$v_+ \lesssim D/L_w \sim 0.15 - 0.3$ (the wall thickness $L_w \sim (15-40)/T$ and the diffusion 
constant $D \sim 5/T$ \cite{Joyce:1994fu}).    
     
\section{Departure of $v_{+}$ from $V_{w}$.}

\begin{figure}[ht]
\begin{center}
\includegraphics[width=0.43\textwidth, clip]{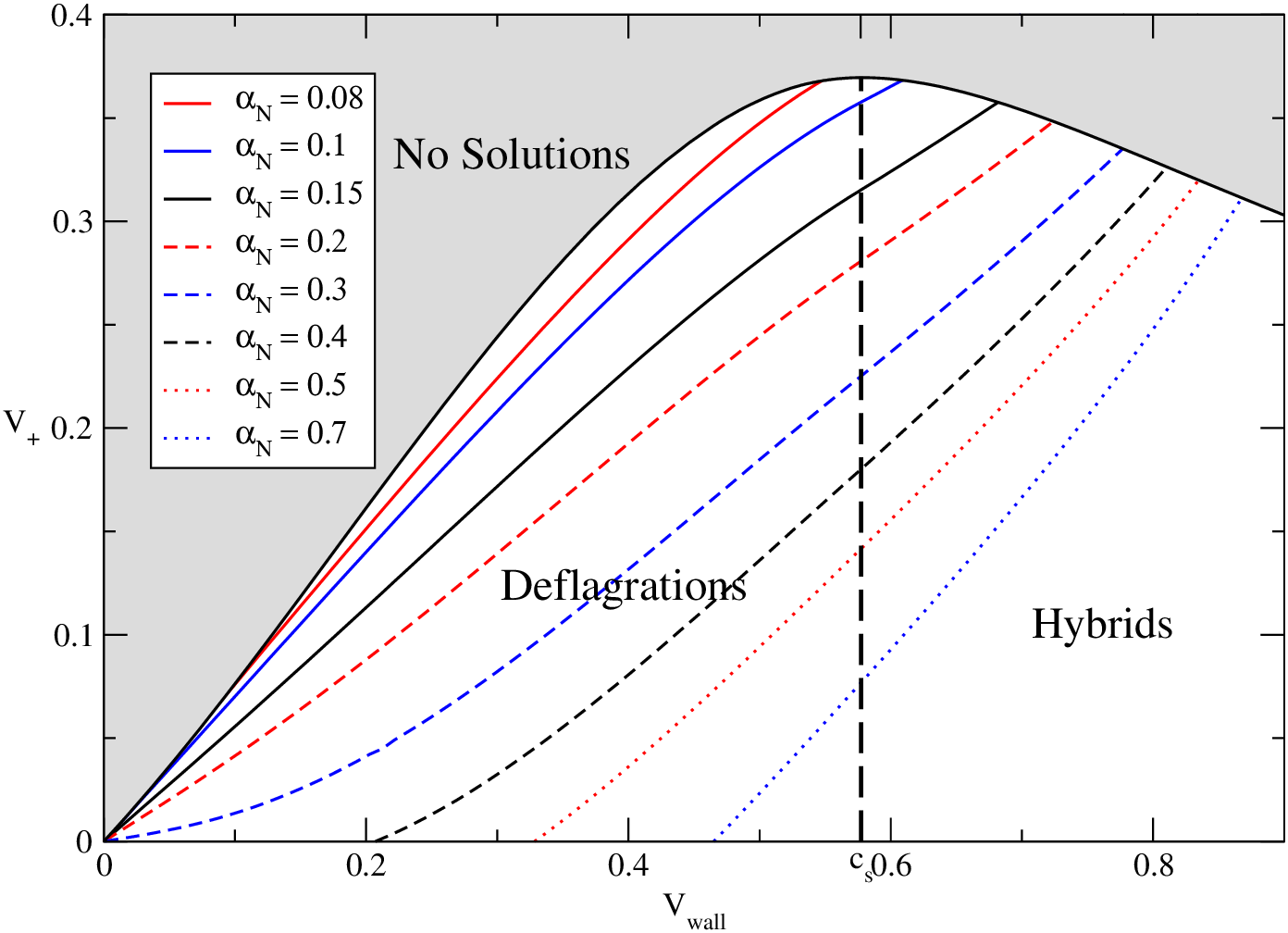}
\caption{\label{fig:2}
\small Relation $v_{+}(\alpha_{N}, V_{w})$.}
\end{center}
\end{figure}

In the presence of a compression wave (deflagrations and hybrid solutions) it is seen 
from (\ref{eq:vvs2}) and (\ref{eq:v_diff}) that $V_w > v_+$. For weak first order phase 
transitions ($\alpha_{N} \ll 1$) one has $v_{+} \simeq V_{w}$ from (\ref{eq:vvs2}) and the effect of considering
$v_{+}$ instead of $V_{w}$ as the relevant velocity for electroweak baryogenesis is small.
However, as $\alpha_{N}$ gets larger (always keeping $\alpha_+ < 1/3$ \cite{EKNS}) and the phase transition gets stronger, 
$v_{+}$ progressively departs from $V_{w}$, eventually reaching $v_{+} \ll V_{w}$ for very strong phase transitions. 
This is shown in Figure \ref{fig:2}, where $v_{+}(V_{w})$ is plotted for increasing values of $\alpha_N$.
Then, for rather strong phase transitions, $v_{+}$ can be kept 
small enough to satisfy the diffusion upper bound for electroweak baryogenesis $v_{+} \lesssim 0.15 - 0.3$  
with a rather large $V_{w}$ ($V_{w} \sim c_{s}$ for deflagrations or even $V_{w} > c_{s}$ for hybrid solutions).

\begin{figure}[ht]
\begin{center}
\includegraphics[width=0.45\textwidth, clip ]{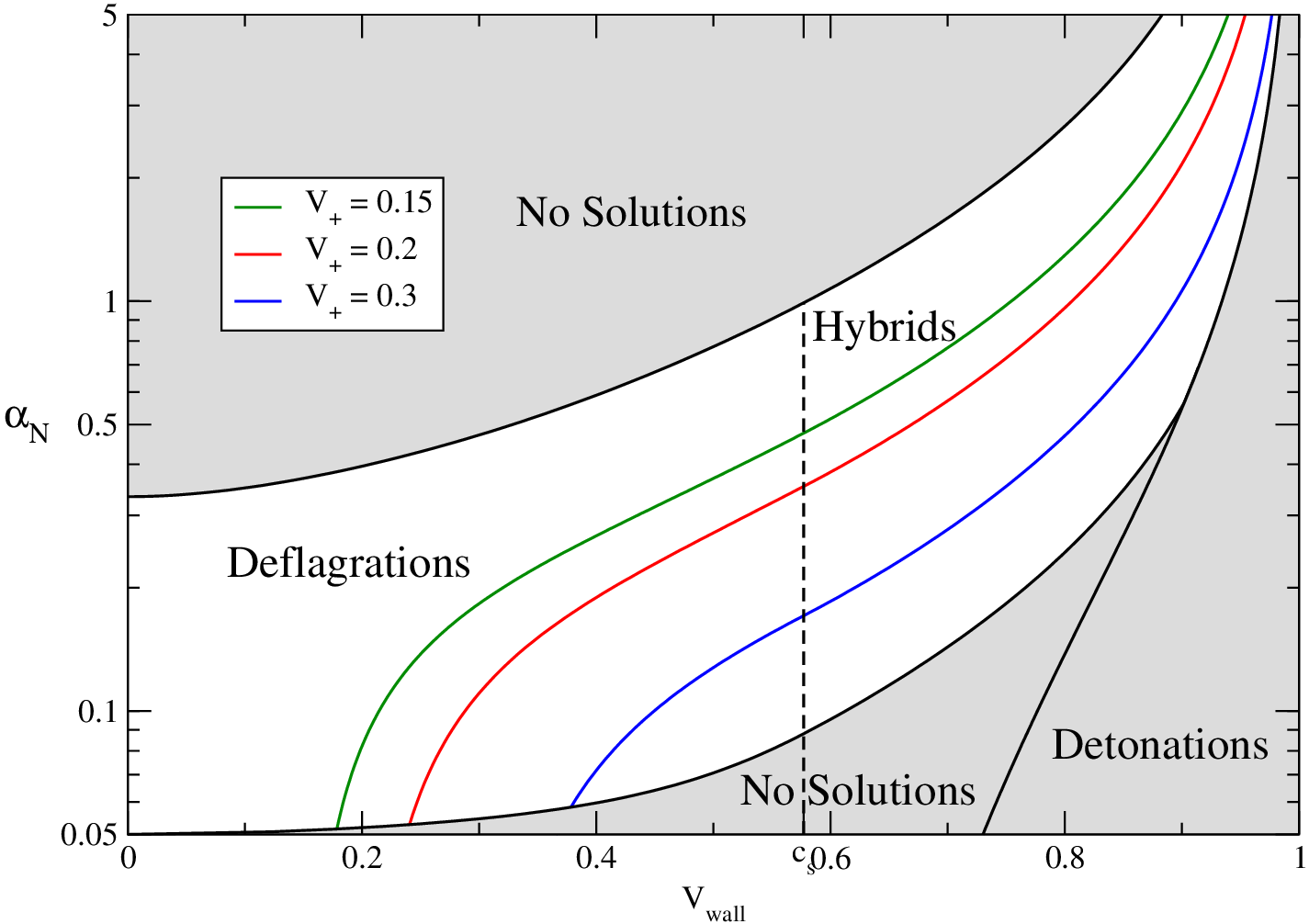}
\caption{\label{fig:3}
\small 
Regions in the ($V_{w}, \alpha_{N}$) plane compatible with $v_{+} < 0.15$, $v_{+} < 0.2$ and $v_{+} < 0.3$
(region above each line).}
\end{center}
\end{figure}

In Figure \ref{fig:3} the region in the parameter space of $\alpha_{N}$ and $V_{w}$ compatible 
with electroweak baryogenesis (for various values of the upper bound on $v_{+}$) is shown. 
For large $\alpha_N$ and small $V_w $, $\alpha_+ > \frac{1}{3}$ and no solutions exist \cite{EKNS}. 
Also, for small $\alpha_N$ there is a maximum value $V_w$ can take with positive plasma friction \cite{KN}, 
and $\alpha_N$ has to be larger than a critical value $\alpha_c$ for bubble expansion to be possible ($\alpha_c \simeq $ 0.05 
for $a_{-}/a_{+} = 0.85$) \cite{EKNS,KN}.  

\section{How Big Can the Gravitational Wave Signal Be?}

Here we will concentrate on production
of gravitational waves through bubble collisions during an electroweak first order phase transition 
(the analysis can be extended to the case of turbulence,
with similar conclusions). The amplitude and peak frequency of the generated stochastic spectrum 
are \cite{Huber:2008hg} (see also \cite{Kamionkowski:1993fg,Caprini:2007xq}):

\begin{small}
\bea
\label{GW_amplitude}
\Omega_{GW}h^2(f_{coll}) \simeq 10^{-6} \left(\frac{100}{g_{*}}\right)^{\frac{1}{3}} 
\left(\frac{H_{*}}{\beta}\right)^2 \quad \quad \quad \quad \quad \quad \nonumber \\
\left(\kappa (\alpha_N,V_w) \right)^2
\left(\frac{\alpha_N}{1+\alpha_N}\right)^2 \frac{1.84 \, V^3_w}{0.42 + V^2_w}
\eea 
\end{small}

\begin{figure}[ht]
\begin{center}
\includegraphics[width=0.48\textwidth, clip ]{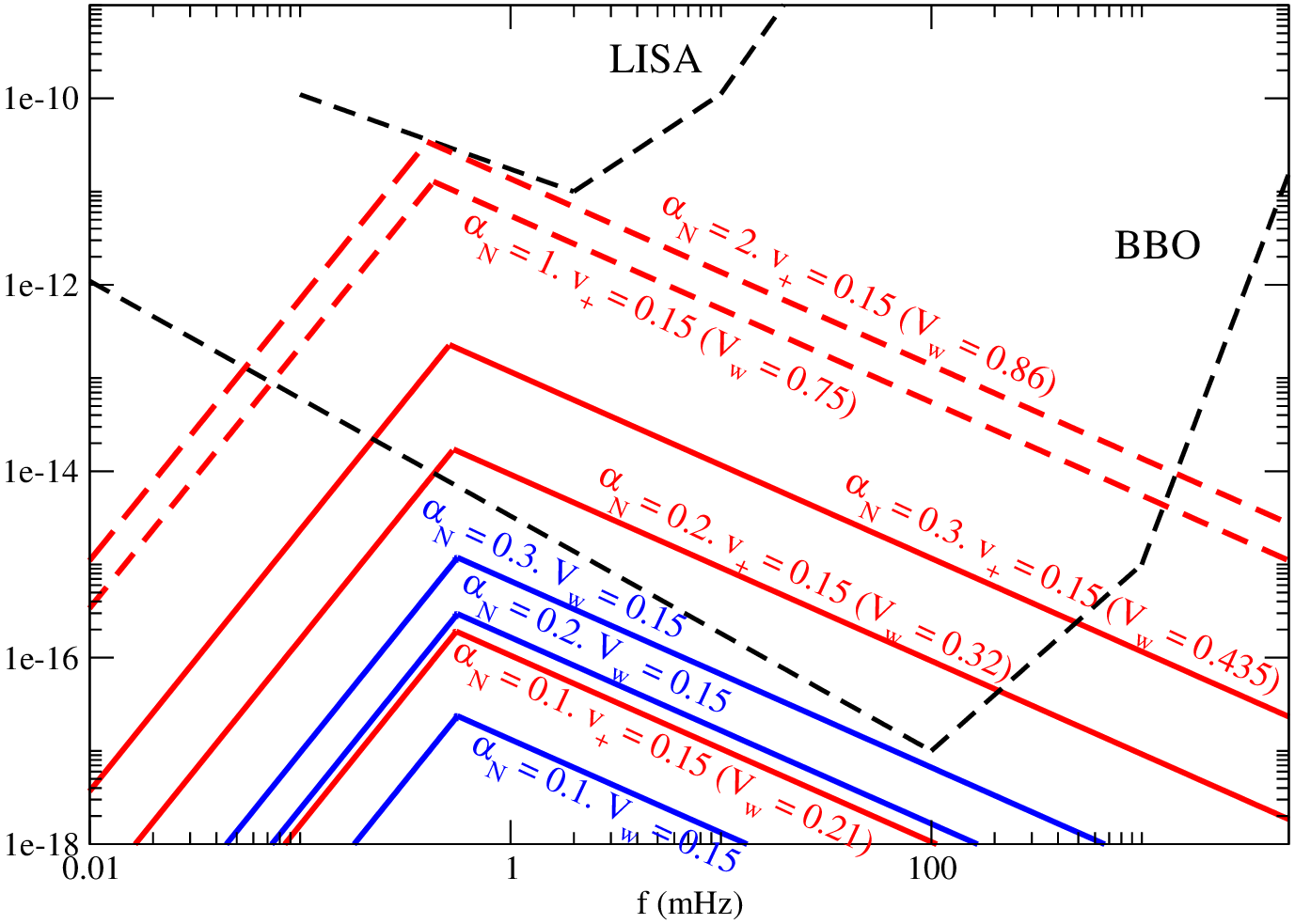}
\caption{\label{fig:4}
\small 
$\Omega_{GW} h^2 (f)$ for various values of $\alpha_N$ and $V_w = 0.15$ (blue lines) or $v_{+} = 0.15$ (red lines).
}
\end{center}
\end{figure}

\begin{small}
\be
\label{GW_frequency}
f_{coll} \simeq 10^{-2} \mathrm{mHz} \left(\frac{g_{*}}{100}\right)^{\frac{1}{6}}  
\left(\frac{T}{100 \, \mathrm{GeV}}\right)^2 \frac{\beta}{H_{*}} \frac{1.2}{1.8 + V^2_w}
\ee
\end{small}

and the spectrum grows as $f^{3}$ for frequencies smaller than $f_{coll}$ \cite{Kamionkowski:1993fg,Caprini:2007xq} 
and falls off as $f^{-1}$ for large frequencies \cite{Huber:2008hg}. 
Typically $\beta/H_{*} \sim 100$ \cite{Hogan:1984hx} and for the electroweak phase transition 
$T \sim 100 \, \mathrm{GeV}$ and $g_{*} \sim 100$. Then, if viable electroweak baryogenesis
would require $V_w < 0.15-0.3$, this would suppress the gravitational wave amplitude both 
explicitly and through the efficiency coefficient $\kappa(\alpha_N,V_w)$ dependence even for 
rather strong phase transitions ($\alpha_N \sim \mathcal{O}(0.1-0.2)$), making the gravitational wave signal in 
scenarios where electroweak baryogenesis is possible undetectable
both at LISA and BBO (blue lines in Figure \ref{fig:4}). Moreover, for very strong transitions ($\alpha_N \sim \mathcal{O}(1)$)
it would be impossible to satisfy the bound $V_w < 0.15-0.3$ and electroweak baryogenesis would simply not be possible. 

However, since the electroweak
baryogenesis bound actually applies to $v_+$, the gravitational wave signal amplitude for a scenario where the bound is satisfied
gets enhanced by 2-3 orders of magnitude with respect to the previous situation (for a transition of the same strength), and it is then
possible to achieve electroweak baryogenesis in scenarios where the gravitational
wave signal may be detectable at BBO for moderately strong phase transitions (solid red lines in Figure \ref{fig:4}) or even marginally 
at LISA for extremely strong transitions (dashed red lines in Figure \ref{fig:4}).

\section{Conclusions.}

If the electroweak phase transition is of first order, 
it will proceed by bubble nucleation and expansion. In 
the presence of a compression wave in front of the expanding bubble walls (which always occurs if the bubbles expand
subsonically and can also occur under certain conditions if the bubbles expand supersonically), the relative velocity
between the wall and the plasma in front ($v_+$) is smaller than $V_w$. While this effect is small 
for weak phase transitions (for which one gets $v_{+} \simeq V_{w}$), it becomes important as the phase transition gets
stronger, and for rather strong transitions one has $v_{+} \ll V_{w}$. Since the background velocity of the plasma 
relevant for the electroweak baryogenesis process is $v_+$ whereas the relevant velocity for gravitational wave production 
through bubble collisions is $V_w$ (see however \cite{Caprini:2007xq}), 
this opens the possibility of achieving viable electroweak baryogenesis (satisfying the diffusion 
bound $v_+ \lesssim 0.15 - 0.3$) and a sizable gravitational wave signal in the same scenario, which was previously regarded
as not possible due to the very different ranges of wall velocities $V_w$ that were thought to be required for 
each the two processes to be efficient.
We actually find that in electroweak baryogenesis scenarios, the gravitational wave signal 
can be easily detected by BBO for moderately strong phase transitions ($\alpha_N \sim \mathcal{O}(0.2-0.3)$), 
and is very close to the LISA sensitivity curve for extremely strong transitions ($\alpha_N \sim \mathcal{O}(1)$). 

However,
whereas a moderately strong first order phase transition is a natural possibility in many theories beyond the Standard Model 
(and so a positive signal at BBO is plausible in these scenarios), the occurence of such a strong transition as to observe a signal with 
LISA is quite unlikely since it corresponds to a very fine-tuned scenario, and also 
in this last case the bubble expansion would have to proceed through hybrids ($V_w > c_s$)
and it has been argued that hybrids are not in general stable solutions, but may evolve into detonations \cite{Laine:1993ey}. 
One should have all these issues in mind when dealing with a concrete model. 

\begin{acknowledgments}
I thank T. Konstandin, J. R. Espinosa and G. Ballesteros for very useful discussions. 
Work supported by the European Commission under contract
PITN-GA-2009-237920 and by the Agence Nationale de la Recherche. 
\end{acknowledgments}


\begin{thebibliography}{99}
\vspace*{-1cm}
%
\bibitem{Cohen:1990py}
  A.~G.~Cohen, D.~B.~Kaplan and A.~E.~Nelson,
  Phys.\ Lett.\  B {\bf 245} (1990) 561;
  Nucl.\ Phys.\  B {\bf 349} (1991) 727;
  Nucl.\ Phys.\  B {\bf 373} (1992) 453;
  Phys.\ Lett.\  B {\bf 336} (1994) 41. 
  
\bibitem{Joyce:1994fu}
  M.~Joyce, T.~Prokopec and N.~Turok,
  Phys.\ Rev.\ Lett.\  {\bf 75} (1995) 1695
  [Erratum-ibid.\  {\bf 75} (1995) 3375];
  Phys.\ Rev.\  D {\bf 53} (1996) 2930;
  Phys.\ Rev.\  D {\bf 53} (1996) 2958.


\bibitem{Witten:1984rs}
  E.~Witten,
  Phys.\ Rev.\  D {\bf 30}, 272 (1984).

\bibitem{Kosowsky:1991ua}
  A.~Kosowsky, M.~S.~Turner and R.~Watkins,
  Phys.\ Rev.\  D {\bf 45} (1992) 4514;
  Phys.\ Rev.\ Lett.\  {\bf 69}, 2026 (1992);
  A.~Kosowsky and M.~S.~Turner,
  Phys.\ Rev.\  D {\bf 47} (1993) 4372.

\bibitem{Kamionkowski:1993fg}
  M.~Kamionkowski, A.~Kosowsky and M.~S.~Turner,
  Phys.\ Rev.\  D {\bf 49} (1994) 2837.


\bibitem{Landau}
  L.~D.~Landau and E.~M.~Lifshitz,
  ``Fluid Mechanics,''
  Pergamon Press, New York, 1989.

\bibitem{Steinhardt:1981ct}
  P.~J.~Steinhardt,
  Phys.\ Rev.\  D {\bf 25} (1982) 2074.

\bibitem{Moore:1995si}
  G.~D.~Moore and T.~Prokopec,
  Phys.\ Rev.\  D {\bf 52} (1995) 7182;

\bibitem{Ignatius:1993qn}
  J.~Ignatius, K.~Kajantie, H.~Kurki-Suonio and M.~Laine,
  Phys.\ Rev.\  D {\bf 49} (1994) 3854.



\bibitem{Caprini:2007xq}
  C.~Caprini, R.~Durrer and G.~Servant,
  Phys.\ Rev.\  D {\bf 77} (2008) 124015.

\bibitem{Huber:2008hg}
  S.~J.~Huber and T.~Konstandin,
  JCAP {\bf 0809} (2008) 022.

\bibitem{Caprini:2009fx}
  C.~Caprini, R.~Durrer, T.~Konstandin and G.~Servant,
  Phys.\ Rev.\  {\bf D79 } (2009)  083519.


\bibitem{Kosowsky:2001xp}
  A.~Kosowsky, A.~Mack and T.~Kahniashvili,
  Phys.\ Rev.\  D {\bf 66} (2002) 024030;
  A.~D.~Dolgov, D.~Grasso and A.~Nicolis,
  Phys.\ Rev.\  D {\bf 66} (2002) 103505;
  G.~Gogoberidze, T.~Kahniashvili and A.~Kosowsky,
  Phys.\ Rev.\  D {\bf 76} (2007) 083002.



\bibitem{Caprini:2006jb}
  C.~Caprini and R.~Durrer,
  Phys.\ Rev.\  D {\bf 74} (2006) 063521;
  C.~Caprini, R.~Durrer and G.~Servant,
  JCAP {\bf 0912} (2009) 024.



\bibitem{EKNS}
  J.~R.~Espinosa, T.~Konstandin, J.~M.~No and G.~Servant,
  JCAP {\bf 1006} (2010) 028.

\bibitem{cite:fate}
  S.~R.~Coleman,
  Phys.\ Rev.\  D {\bf 15}, 2929 (1977)
  [Erratum-ibid.\  D {\bf 16}, 1248 (1977)];
  C.~G.~Callan and S.~R.~Coleman,
  Phys.\ Rev.\  D {\bf 16}, 1762 (1977);
  A.~D.~Linde,
  Phys.\ Lett.\  B {\bf 100}, 37 (1981).

\bibitem{Laine:1993ey}
  M.~Laine,
  Phys.\ Rev.\  D {\bf 49} (1994) 3847;
  H.~Kurki-Suonio and M.~Laine,
  Phys.\ Rev.\  D {\bf 51}, 5431 (1995).

\bibitem{KN}
  T.~Konstandin and J.~M.~No,
  JCAP {\bf 1102} (2011) 008

\bibitem{Hogan:1984hx}
  C.~J.~Hogan,
  Phys.\ Lett.\  B {\bf 133} (1983) 172.

\bibitem{LEP}
  R.~Barate {\it et al.} [ LEP Working Group for Higgs boson searches and ALEPH and DELPHI and L3 and OPAL Collaborations ],
  Phys.\ Lett.\  {\bf B565 } (2003)  61-75.

\bibitem{Kajantie}
  K.~Kajantie, M.~Laine, K.~Rummukainen, M.~E.~Shaposhnikov,
  Nucl.\ Phys.\  {\bf B466 } (1996)  189-258; 
  Phys.\ Rev.\ Lett.\  {\bf 77 } (1996)  2887-2890; 
  Nucl.\ Phys.\  {\bf B493 } (1997)  413-438.

\bibitem{Pietroni:1992in}
  M.~Pietroni,
  Nucl.\ Phys.\  {\bf B402 } (1993)  27-45.

\bibitem{KonstandinNMSSM}
  S.~J.~Huber and T.~Konstandin,
  JCAP {\bf 0805} (2008) 017

\bibitem{Blum}
  K.~Blum, C.~Delaunay and Y.~Hochberg,
  Phys.\ Rev.\  D {\bf 80} (2009) 075004

\bibitem{Germano}
  M.~Carena, G.~Nardini, M.~Quiros and C.~E.~M.~Wagner,
  Nucl.\ Phys.\  B {\bf 812} (2009) 243

\bibitem{Bochkarev:1990fx}
  A.~I.~Bochkarev, S.~V.~Kuzmin, M.~E.~Shaposhnikov,
  Phys.\ Lett.\  {\bf B244 } (1990)  275-278.

\bibitem{Turok:1991uc}
  N.~Turok, J.~Zadrozny,
  Nucl.\ Phys.\  {\bf B369 } (1992)  729-742.

\bibitem{Davies:1994id}
  A.~T.~Davies, C.~D.~froggatt, G.~Jenkins, R.~G.~Moorhouse,
  Phys.\ Lett.\  {\bf B336 } (1994)  464-470.

\bibitem{Cline:1996mga}
  J.~M.~Cline, P.~-A.~Lemieux,
  Phys.\ Rev.\  {\bf D55 } (1997)  3873-3881.

\bibitem{AndersonHall}
  G.~W.~Anderson, L.~J.~Hall,
  Phys.\ Rev.\  {\bf D45 } (1992)  2685-2698.

\bibitem{Ham:2004cf}
  S.~W.~Ham, Y.~S.~Jeong, S.~K.~Oh,
  J.\ Phys.\ G {\bf G31 } (2005)  857-872.

\bibitem{Espinosa:2007qk}
  J.~R.~Espinosa, M.~Quiros,
  Phys.\ Rev.\  {\bf D76 } (2007)  076004.

\bibitem{Profumo:2007wc}
  S.~Profumo, M.~J.~Ramsey-Musolf, G.~Shaughnessy,
  JHEP {\bf 0708 } (2007)  010.

\bibitem{EKNQ}
  J.~R.~Espinosa, T.~Konstandin, J.~M.~No, M.~Quiros,
  Phys.\ Rev.\  {\bf D78 } (2008)  123528.


\bibitem{Servant}
  C.~Grojean, G.~Servant and J.~D.~Wells,
  Phys.\ Rev.\  D {\bf 71} (2005) 036001; 
  C.~Delaunay, C.~Grojean and J.~D.~Wells,
  JHEP {\bf 0804} (2008) 029

\bibitem{Friction}
  P.~John and M.~G.~Schmidt,
  Nucl.\ Phys.\  B {\bf 598} (2001) 291
  [Erratum-ibid.\  B {\bf 648} (2003) 449].

\end{thebibliography}
\end{document}